\documentclass[11pt,a4paper]{amsart}

\usepackage[latin1]{inputenc}
\usepackage[english]{babel}
\usepackage{amsmath}

\usepackage{amssymb, mathabx}
\usepackage[numbers]{natbib}
\usepackage{graphicx}
\usepackage{braket}
\usepackage[pdftex,plainpages=false,colorlinks,hyperindex,bookmarksopen,linkcolor=red,citecolor=blue,urlcolor=blue]{hyperref}

\usepackage{epstopdf}

\usepackage{braket}

\bibpunct{[}{]}{;}{n}{,}{,}
\usepackage[hmargin=3cm,vmargin={3.5cm,4cm}]{geometry}

\theoremstyle{definition}                                 

\theoremstyle{definition}                           

\theoremstyle{remark}                             

\usepackage{color}

\newcommand{\R}{\mathbb{R}}  
\newcommand{\C}{\mathbb{C}} 
\newcommand{\N}{\mathbb{N}} 
\def\eg{{\it e.g. }} 
\def\ie{{\it i.e. }}

\newcommand{\res}{\mathcal{R}es \,}
\newcommand{\B}[1]{{\color{black}#1}}

\def\d{\partial}

\def\ds{\displaystyle}

\numberwithin{equation}{section}

\allowdisplaybreaks

\begin{document}
\title{A dynamic  viscoelastic analogy	\\ for  fluid-filled elastic tubes}

    \author{Andrea Giusti$^1$}
		\address{${}^1$ Department of Physics $\&$ Astronomy, University of 	
    	    Bologna and INFN. Via Irnerio 46, Bologna, ITALY.}
		\email{andrea.giusti@bo.infn.it}
	
    \author{Francesco Mainardi$^2$}
    	    \address{${}^2$ Department of Physics $\&$ Astronomy, University of 	
    	    Bologna and INFN. Via Irnerio 46, Bologna, ITALY.}
			\email{francesco.mainardi@bo.infn.it}
			

    \date{January 2016}

\begin{abstract}
In this paper we evaluate  the dynamic effects of the fluid viscosity 
      for fluid filled elastic tubes in the framework of a linear uni-axial theory.
      Because of the linear approximation,  the effects on  the fluid inside the elastic tube are taken into account according to the Womersley theory for a pulsatile flow in a rigid tube.
      The evolution \B{equations} for \B{the response variables are} derived by means of the Laplace transform technique and \B{they all turn} out to be \B{the very same} integro-differential equation of the convolution type. This equation has the same structure as the one describing uni-axial waves in linear viscoelastic solids characterized by  a relaxation modulus or by a creep compliance. 
In our case, the analogy is connected with a peculiar viscoelastic solid which exhibits creep properties similar to those of a fractional Maxwell model (of order $1/2$) for short times, and of a standard Maxwell model for long times.
       The present analysis could find applications in biophysics concerning the propagation of pressure waves within large arteries.
\end{abstract}

    \maketitle
    
    \section{Introduction} \label{section_intro}
    
    It is well known that the dynamic theory for fluid filled elastic tubes
    \B{might} find applications \B{in the analysis of} the propagation of pressure waves within
large arteries. 
In the framework of a uni-axial theory, the \B{effect of the viscosity of} the fluid inside the tube (\eg  blood) is usually taken into account \B{including a friction term in}
the one-dimensional momentum equation, due to the \B{Poiseuille's formula for steady flows in rigid tubes}, see \eg \cite{Hoogstraten-Smit 1978}, and \cite{BMM 1981}.\\
Restricting ourselves to a linear approach, here  we find it more appropriate to evaluate the friction term \B{for} a non-stationary flow, which is easily derived starting from a sinusoidal flow, \B{as in the renowned Womersley model}
for pulsatile flow of viscous (Newtonian) fluids in rigid tubes \cite{Womersley 1955}, see also \cite{Daidzic JFE2014}, \cite{He-et-al ABE1993}.

In Section 2 we follow the approach displayed in \cite{Varley-et-al 1966} considering the Navier-Stokes equations for an incompressible Newtonian fluid in cylindrical coordinates. \B{For such system of PDEs} we neglect the motion in the circumferential direction \B{and we assume the flow to be quasi-one-dimensional}. \B{In particular, the radial velocity is assumed to be very small with respect to the axial one}.
The resulting equations are then integrated over the radial coordinate, reducing the number of independent variables to two (\ie time $t$ and axial distance $x$) and the dependent variables to three (pressure, averaged velocity, and cross-sectional area).
A pressure-area relation for a uniform elastic tube, longitudinally
tethered, is then used to eliminate one of the \B{remaining} dependent variables.
However, the final equations contain two quantities which depend on the velocity profile.
 
In Section 3, \B{working within} the framework of a linear uni-axial theory, 
\B{we reduce the system of PDEs discussed in Section 2 to a system} of two linear equations that contains only an additional unknown function, \ie the friction term, that we denote by $f_0(x,t)$ 
and which still depends on the velocity profile.
\B{Then}, because of our linearity assumption,
 we find it reasonable to evaluate $f_0$ in terms of the averaged velocity. This can be done 
 by assuming the velocity profile to be described as in the case of an unsteady flow in a rigid tube with an unperturbed radius based on the Womersley theory for pulsatile flow.
As a consequence, we are able to obtain the friction term corresponding to any time history of the pressure-gradient 
by means of a time convolution integral between a certain memory function $\Phi(t)$ and the pressure-gradient itself.
Finally, we obtain the evolution equation for the pressure waves \B{that appears to have} the same structure \B{as the one known} for stress waves in a viscoelastic solid (in the relaxation representation), see \eg \cite{Mainardi 2010}. 

In Section 4 we \B{discuss} in detail this dynamic analogy in the framework of a linear theory of viscoelasticity. \B{For further details about mathematical and historical aspects of this theory we recommend the interested reader to refer to specialized articles and treatises, \eg \cite{Rabotnov-1980}, \cite{K-2010}, \cite{Mainardi 2010}, \cite{MR-2014} and \cite{Getal-2014}.}\\
The memory function $\Phi(t)$ \B{is then} shown to be a complete monotonic function expressed in terms of a Dirichlet series. \B{Therefore, the latter has to be intended} as the rate of the relaxation modulus of a peculiar viscoelastic model.
We then calculate the corresponding rate of creep $\Psi(t)$, 
which enables us to obtain the wave equation in the creep representation.
Finally, by a suitable time integration \B{of $\Phi(t)$ and $\Psi(t)$}, we derive the corresponding relaxation modulus $G(t)$ and the creep compliance $J(t)$, providing a full characterization of the viscoelastic analogy.

In Section 5 we derive the asymptotic representations of the memory functions $\Phi(t)$
and $\Psi(t) $ for short and long times.
The resulting functions show that the \B{corresponding} viscoelastic model is governed by a stress-strain relation of Maxwell type with a fractional order changing  from 1/2  to 1
as time evolves from zero to infinity.
 For sake of clarity, we show the plots versus time of the functions $\Phi(t), \Psi(t$) and of $G(t)$ and $J(t)$.
 
Finally, in Section 6  we complete the paper with concluding remarks and hints for future research.  Furthermore, a detailed mathematical description of the technical results can be found in the Appendix. 
 
\section{The basic non-linear equations} \label{section_2}

Let us consider in a cylindrical coordinate system
 the Navier-Stokes equations of motion for an incompressible Newtonian fluid. 
 If the motion in the circumferential direction is
neglected, these equations read (denoting the partial derivatives with subscripts)
\begin{equation} \label{eq_2-1}
\left\{
\begin{split}
&u_{t} + w u_{r} + u u_{x} + \frac{p_{x}}{\rho} = \nu \left( u_{rr} + \frac{u_{r}}{r} + u_{xx} \right)\,,\\
&w_{t} + w w_{r} + u w_{x} + \frac{p_{r}}{\rho} = \nu \left( w_{rr} + \frac{w_{r}}{r} - \frac{w}{r^2} + w_{xx} \right)\,, 
\end{split}
\right.
\end{equation}
where $t$ is the time, $x$, $r$ denote the axial and radial directions, $u$, $w$ the corresponding components of the fluid velocity and $p$ the pressure.  
The constants $\rho$ and $\nu$ are the density and kinematic viscosity of the fluid.

Then, following the approach  shown in \cite{Varley-et-al 1966}, the flow is assumed to be quasi one dimensional, \ie the radial velocity $w$ is very small with respect to the axial velocity $u$ 
and the resulting equations are integrated over the radial coordinate $r$.
As a consequence we arrive at  Eqs. (14) and (15) in \cite{Varley-et-al 1966}, which in our notation read:


\begin{equation}\label{eq_2-2}
\left \{
\begin{split}
&U_t + \frac{U}{A} (1-\chi) A_t + \chi UU_x + \frac{p_x}{\rho} =
 \left. \frac{2 \nu}{R} u_r \right| _R\,, \\
&A_t + \frac{\d}{\d x} \left(U A \right) = 0\,.
\end{split}
\right .
\end{equation} 
In these equations
\begin{equation} \label{eq_2-3}
U = {\ds \frac{1}{R^2}}\, {\ds \int_0^R \!\! 2ru\, dr}
\end{equation}
 is the averaged (uni-axial) velocity (the quantity measured in most experiments),  
\begin{equation} \label{eq_2-4}
A= \pi R^2
\end{equation}
 is the cross-sectional area of the tube, and
\begin{equation} \label{eq_2-5}
\chi = {\ds \frac{1}{R^2\,U^2}}\, {\ds \int_0^R \!\! 2 r u^2\, dr}\,.
\end{equation}
As a consequence,  $\{ U(x,t),  A(x,t)\}$, 
the basic dependent variables for a quasi-one dimensional flow, 
turn out to be related \B{also} to the pressure gradient $\Lambda =p_x$,
 to the parameter $\chi$
and to the friction term defined by 
\begin{equation} \label{eq_2-6}
f(x,t) =    \left. \frac{2 \nu}{R} u_r \right| _{R} \, .
\end{equation}
As a matter of fact,  both quantities depend on the variation of the axial velocity  $u$ with $r$ (the velocity profile) so they  could not  be     properly determined without  carrying out a full two-dimensional non-linear  treatment of the problem.
\B{It is worth stressing that when} the fluid is inviscid the velocity profile is flat ($\chi= 1$) and the friction term is absent ($f(x,t) \equiv 0$). 

According to  \cite{Varley-et-al 1966}, in the presence of viscosity, a particularly simple assumption is the parabolic form of the velocity profile:
 \begin{equation} \label{eq_2-7}
u= 2U(1-r^2/R^2) \, ,
\end{equation}
so that  $\alpha = 4/3$ and 
\begin{equation} \label{eq_2-8}
f(x,t) = - 8 \nu U/R^2\,.
\end{equation}
This assumption, however, is too simple because it is satisfied only for stationary flows related to the Poiseuille approximation.

  \section{The basic linear equations} \label{section_3}

Linearizing  Eqs.  (\ref{eq_2-2})  
we get the following systems of PDEs:
\begin{equation} \label{eq_3-1}
\left\{
\begin{aligned}
&U_t + p_x / \rho = f_0 (x, t)\,, \\
&A_t + A_0 U_x = 0 \,,
\end{aligned}
\right. 
\end{equation}
from which we get
\begin{equation} \label{eq_3-2}
\left\{
\begin{aligned}
&A_0 U_t + c_0 ^2 A_x = A_0 \,f_0 (x, t)\,, \\
&A_t + A_0 U_x = 0 \,,\\
\end{aligned}
\right.
\end{equation}
where
 $A_0= \pi R_0 ^2$ is the unperturbed cross-sectional area of radius $R_0$,
\begin{equation}  \label{eq_3-3}
 c_0^2 = \frac{A_0}{\rho}\, \left.\left(\frac{dp}{dA}\right) \right|_0
 \end{equation}
 denotes the corresponding Moens-Korteweg velocity
(for pressure waves in elastic tubes filled with an inviscid fluid) 
 and
 \begin{equation}  \label{eq_3-4}
 f_0 =  \left. \frac{2 \nu}{R_0} u_r \right| _{R_0}
\end{equation}
is the corresponding friction term due to the viscosity,  evaluated at the same unperturbed radius.

In the case of a parabolic velocity profile (Poiseuille approximation of steady flow), \B{linearizing Eqs.  (\ref{eq_2-7}) and (\ref{eq_2-8})}, we get 
\begin{equation}\label{eq_3-5}
\begin{split}
\qquad u &= 2 U(x,t)\,\left (1-r^2/R_0^2 \right) \\
\Rightarrow \quad &f_0(x,t) = -\frac{8 U(x,t)}{\tau} \,, \; \tau = \frac{R_0 ^2}{\nu}\,.
\end{split}
\end{equation}
We note that the time constant  $\tau$, introduced here for steady flow, would provide a proper time scale also in all  profiles occurring for any unsteady flow.
 As a consequence, in every function of time appearing in our approach  the constant
  $\tau$ is to be interpreted as a proper time scale introduced by the viscosity of the fluid; in future, even when we will omit 
  $\tau$ in the time dependence,  this scale-parameter is understood to be equal to one.

In order to eliminate   the velocity component in the radial direction, in the framework of a linearized  theory for unsteady flow, and evaluate  the corresponding friction term $f_0$ in terms of the uni-axial velocity $U(x,t)$, we find it reasonable 
to take profit of the well established \B{Womersley linear model}
for pulsatile flow of a viscous (Newtonian) fluid in rigid tubes \cite{Womersley 1955}.  

Because of the linearity, we can easily compute the friction term corresponding
to a \B{pressure-gradient which is sinusoidal in time and oscillates with a frequency} $\omega$, 
$$\widehat  \lambda(x,t;\omega) = \Lambda_0 (x)\, \exp \left(- i \omega t \right)$$
as derived by \B{the Womersley model}.
Introducing the Womersley parameter\footnote{In our analysis  we have preferred
to express the Womersley parameter $\alpha$   in terms of the time scale $\tau$  
introduced in Eq.   (\ref{eq_3-5}).} 
\begin{equation}\label{eq_3-6}
\alpha := R_0 \sqrt{\frac{\omega}{\nu}} = \sqrt{\omega \tau} \, ,
\end{equation}
\B{we get the friction expressed in terms of} $(\sqrt{\omega \tau})$:
\begin{equation}\label{eq_3-7}
\widehat{f_0} (x,t;\omega) = \widehat{\Phi} (\omega) \, \Lambda_0 (x) \, 
\exp(-i\omega t) / \rho\,,
\end{equation}
with
\begin{equation}\label{eq_3-8}
\widehat{\Phi} (\omega) =\frac{2 }{i^{3/2}\sqrt{\omega\tau }} \,
\frac{ J_1(i^{3/2}\sqrt{\omega\tau })}{  J_0(i^{3/2}\sqrt{\omega\tau })}\,,
\end{equation}
where  $J_0$, $J_1$ denote the   Bessel functions of order $0$, $1$, respectively, see e.g. \cite{AS 1965}.

Then, being interested  to transient motion,  it is worth to pass from Fourier transform to Laplace transform with parameter  $s= i\omega$ so that we get    
\begin{equation}\label{eq_3-9}
\widetilde{\Phi} (s ) = \frac{2} {\sqrt{s\tau }} \,
\frac{I_1(\sqrt{s\tau })} { I_0(\sqrt{s\tau })}\,,
\end{equation}
where  $I_0$, $I_1$ denote the modified Bessel functions of order $0$, $1$, respectively.  We have used the known relation between Bessel and modified Bessel functions of any  order $\nu$, $ J_\nu(iz) = i^\nu I_\nu(z) $, see  e.g. \cite{AS 1965}.

As a consequence we
can formally obtain the friction term corresponding to any time history of the pressure-gradient $\Lambda (x,t)$ by means of a Laplace convolution integral.
Indeed we get:
\begin{equation} \label{eq_3-10}
f_0 (x, t) =  \Phi (t) \ast \Lambda (x, t) / \rho
\end{equation}
from which we derive
\begin{equation} \label{eq_3-11}
U_t (x,t)= - [1 - \Phi (t) \, \ast] \, \Lambda (x,t) / \rho
\end{equation}
where $\Phi (t)$ denotes the inverse Laplace Transform of $\widetilde{\Phi} (s)$
and $\ast$ denotes the time Laplace convolution\footnote{We recall the definition of the time convolution between two locally integrable functions $f(t)$, $g(t)$: $$ f(t) \ast  g (t) := \int _0 ^t f(\tau) g(t - \tau) \, d \tau 
= \int _0 ^t f (t - \tau) g (\tau) \, d \tau \,,$$
so that its Laplace transform reads $\widetilde f(s) \, \widetilde g(s)$.}.\\
	
	So doing we have properly modified  the relation between acceleration $U_t$ and pressure gradient $\Lambda$ for inviscid flow (the Euler equation) in terms of a Laplace convolution with a characteristic function $\Phi(t)$ in order to take into account the \textit{memory effects of the viscosity}.
We agree to call $\Phi(t)$ the {\it relaxation memory function},  whose meaning will be clarified in the following with respect to our wave process.

Indeed, \B{denoting with $Y=Y (x,t)$ a generic response variable (such as $\{ U, \, A,\, p, \Lambda \}$), after simple manipulations we get the evolution equation}
\begin{equation}\label{eq_3-12}
Y_{tt}(x,t)  = c_0^2 \, [1- \Phi (t) \, \ast ] \, Y_{xx}(x,t)
\end{equation} 
which, in the absence of viscosity ($f_0(x, t)\equiv 0$,  {\it i.e.}  $\Phi (t) \equiv 0$),
reduces to the classical D'Alembert wave equation
\begin{equation}
Y_{tt}(x,t)  = c_0^2 \,  Y_{xx}(x,t) \, .
\end{equation}

	As a matter of fact, Eq. (\ref{eq_3-12}) is an integro-partial differential equation, which could be  taken into account in order to investigate wave processes in fluid filled compliant tubes
of biophysical interest. 
In particular, this evolution equation	 
	could be solved for $x \ge 0$,  $t \ge 0$ together with a known input condition $Y(0,t) = Y_0 (t)$ at $x=0$, assuming that the tube is initially quiescent, and that there are no waves coming from $x=+\infty$. This initial boundary-value problem is usually referred to as a \textit{signalling problem}.
The above simplifying assumptions are expected to explain the qualitative features of the distortion (due to the viscosity of the fluid) 
 of the transient waves generated at the accessible end of the tube as they propagate along the tube. 
 
  \section{The dynamic  viscoelastic analogy} \label{section_4}

From Eq. (\ref{eq_3-12}) we easily recognize an analogy with the wave equation of linear viscoelasticity in the \textit{Relaxation Representation}, see \cite{Mainardi 2010}.
For this purpose  we will prove that the memory function $\Phi(t)$ is  completely monotonic 
(\ie it is a non-negative, non-increasing  function  with infinitely many derivatives alternating in sign for $t\ge 0$),  as required to represent (with opposite sign) the non-dimensional rate of
the relaxation modulus $G(t)$ scaled with its initial value $0<G(0^+)<\infty $ according to
\begin{equation} \label{eq_4-1}
\Phi(t) = - \frac{1}{G(0^+)} \, \frac{dG}{dt}\,,
\end{equation}
and consequently in Laplace domain
\begin{equation} \label{eq_4-2}
s \widetilde G(s) = G(0^+) \left[ 1- \widetilde \Phi (s) \right]\,.
\end{equation}
 We recall that $G(t)$  represents the stress response to a unit step of strain, see \cite{Mainardi 2010} and it is usually assumed to be completely monotone as well. 
Being  $G(0^+)$ finite and positive, 
the corresponding  viscoelastic solid exhibits stress waves with a finite wave front velocity $c_0= \sqrt{G(0^+)/\rho}$.    

In fact, carrying out the inversion of the Laplace transform (\ref{eq_3-9}),  by means of the Bromwich formula and  applying the residues theorem, we get the memory function in terms of a convergent Dirichlet series as 
\begin{equation} \label{eq_4-3}
\Phi (t) = \frac{4}{\tau} \sum _{n=1} ^\infty \exp \left( - \lambda _n ^2 t / \tau \right)
\end{equation}
where $\lambda_n$ are the zeros of  the oscillating  Bessel function $J_0$
on the positive real axis.
\B{Further details about the proof can be found in the} Appendix. 
\B{So, we} recognize that the relaxation memory function, resulting from the sum of a convergent Dirichlet series with positive coefficients, is indeed a complete monotone function\B{, ensuring} the complete monotonicity of the corresponding relaxation modulus $G(t)$, with $G(+\infty) = 0$.

   In Linear Viscoelasticity, the Relaxation Modulus $G(t)$ is coupled with \B{another} material function, the {\it Creep Compliance} $J(t)$, which represents the strain response to a unit step of stress. 
   \B{Hence}, to complete the dynamic viscoelastic analogy 
we now consider, in addition to the {\it relaxation representation} of the wave equation
(\ref{eq_3-12}), also the corresponding  {\it  creep representation}.  
   
We first observe that the two material functions $G(t)$ and $J(t)$ are connected through their Laplace transforms by the following fundamental equation
\begin{equation} \label{eq_4-4}
s \, \widetilde J(s) = \frac{1}{ s \, \widetilde G(s)}
\end{equation}   
that implies in the time domain
\begin{equation} \label{eq_4-5}
J(0^+) \!=\! \frac{1}{G(0+)}\!>\!0 \,, \,\,\, J(+\infty) \!=\! \frac{1}{G(+\infty)}\!=\!+\infty \,.
\end{equation}
This means, according with the classification of viscoelastic media (see \cite{Mainardi 2010}), the model of our concern can be understood as of Type II.  

Introducing the non-dimensional rate of creep 
$\Psi(t)$, defined as
\begin{equation} \label{eq_4-7}
\Psi(t) := \frac{1}{J(0^+)} \, \frac{dJ}{dt}\,,
\end{equation}  
we have   
\begin{equation} \label{eq_4-8}
s \widetilde J(s) = J(0^+) \left[ 1+ \widetilde \Psi (s) \right]\,.
\end{equation}
  Then,  in the Laplace domain, the  functions $\Phi(t)$ and $\Psi(t)$ turn out to be related as
\begin{equation}\label{eq_4-9}
1 + \widetilde{\Psi} (s) = \left[ 1 - \widetilde{\Phi} (s) \right] ^{-1}
\end{equation}  
as a consequence of Eqs. (\ref{eq_4-2}), (\ref{eq_4-4}) and (\ref{eq_4-8}).

Now, recalling the \textit{recurrence relations} for the modified Bessel functions of the first kind, $I_\nu (z)$, \ie
$$ I_{\nu - 1} (z) - \frac{2 \nu}{z} \, I_{\nu} (z) = I_{\nu + 1} (z)\,,$$
with $\nu =1$,
after a simple manipulation we get:
\begin{equation}\label{eq_4-10}
\widetilde{\Psi} (s) = 
\frac{2}{\sqrt{s\tau}} \, \frac{I_1 (\sqrt{s\tau})}{I_2 (\sqrt{s\tau})}\,.
\end{equation}
	The inversion of the Laplace transform can be carried out using again the Bromwich formula. Then, applying the residues theorem we get: 
\begin{equation} \label{eq_4-11}
\Psi (t) = \frac{8}{\tau} + 
\frac{4}{\tau} \sum _{n=1} ^\infty \exp \left( - \mu_n ^2 t / \tau \right)\,
\end{equation}
where $\mu_n$ are the zeros of the oscillating Bessel function $J_2$
on the positive real axis.

\B{Therefore}, we recognize that the $\Psi$ function, resulting from the sum of a convergent Dirichlet series with positive coefficients, is also a complete monotone function like $\Phi (t)$. 
As a consequence,  the creep compliance $J(t)$ turns out to be  a Bernstein function (that is a non negative function with a \B{completely monotone} first derivative). 
We note, however, the presence of a \B{positive} constant term \B{added to} the Dirichlet series. \B{This implies that the asymptotic expression for the creep compliance $J(t)$, for long times, contains an extra term which is linear in time. The latter is a typical feature of the classical Maxwell model of viscoelasticity}.

	Using the analogy with the linear dynamical theory of viscoelasticity, the wave equation corresponding to the creep representation reads, see \cite{Mainardi 2010},
\begin{equation} \label{eq_4-12}
[1 + \Psi (t) \, \ast \, ] \, Y_{tt}  = c_0^2 \, Y_{xx}\,,
\end{equation}  
where $\Psi(t)$ can be referred to as the {\it creep memory function}.
	Given the previous results, we can now trivially compute the Relaxation Modulus
	$G(t)$ and the Creep Compliance $J(t)$.

In view of Eqs (\ref{eq_4-1}) and (\ref{eq_4-7}),
 we \B{get} 
$G(t)$ and $J(t)$
\begin{equation} \label{eq_4-13a}
  G(t) = G(0^+)\left[1 - \int _0 ^t \Phi (t') \, dt'\right]\, ,
\end{equation}
\begin{equation} \label{eq_4-13b}
  J(t) = J(0^+)\left[1 + \int _0 ^t \Psi (t') \, dt'\right].
\end{equation}
Integrating term by term we get
\begin{equation} \label{eq_4-14a}
G(t) \!=\! G(0^+) \left\{1 - \! \sum _{n=1} ^{\infty} \! \left[\frac{4}{\lambda _n ^2} - 
\frac{4}{\lambda _n ^2} \, e^{- \lambda _n ^2 t / \tau} \right] \right\},
\end{equation}
\begin{equation} \label{eq_4-14b}
J(t) \!=\! J(0^+) \left\{1 + \frac{8t}{\tau} + \! \sum _{n=1} ^{\infty} \! \left[\frac{4}{\mu _n ^2} - 
\frac{4}{\mu _n ^2} \, e^{- \mu _n ^2 t / \tau} \right] \right\} .
\end{equation}
Due to the results for the the zeros of Bessel functions shown in \cite{Sneddon 1960}, \ie
\begin{equation} \label{series_sum}
\sum _{n=1} ^{\infty} \frac{4}{\lambda _n ^2} = 1\,,
\quad J_0(\lambda_n) =0, \, \forall n \in \N \, ,
\end{equation} 
\begin{equation}
\sum _{n=1} ^{\infty} \frac{4}{\mu _n ^2} = \frac{1}{3}\,,
 \quad J_2(\mu_n) =0, \, \forall n \in \N \, ,
\end{equation}
we get
\begin{equation} \label{eq_4-15a}
G(t) = G(0^+) \, \! \sum _{n=1} ^{\infty} 
\frac{4}{\lambda _n ^2} \exp \left( - \lambda _n ^2 t / \tau \right),
\end{equation}
\begin{equation} \label{eq_4-15b}
J(t) = J(0^+) \!
\left\{\frac{4}{3} + \frac{8 t}{\tau} - \sum _{n=1} ^{\infty} \frac{4}{\mu _n ^2} \exp \left( - \mu _n ^2 t / \tau \right) \right\} .
\end{equation}

\B{As we see from Eq.(\ref{eq_4-14a}), together with the condition $G(+\infty) = 0$, we recover the identity in Eq. (\ref{series_sum}). This result gave us a relevant hint to derive Sneddon's result about series involving zeros of Bessel functions of the first kind, by means of a technique based on the Laplace transform (see \cite{AG-FM-eprint}).}


\section{Asymptotic and Numerical results}  \label{section_5}
	In this section we exhibit the plots (versus time) of the relaxation and creep memory functions as computed taking a suitable number of terms in the corresponding Dirichlet series (\ref{eq_4-3}) and (\ref{eq_4-11}), respectively.
	From our numerical experiments, we found that 100 \B{terms are} surely suitable to get stable and confident results.
	
We also compare these plots with the asymptotic representations of the two functions for short and long time in order to check their matching with the numerical solutions.
According  to the Tauberian theorems, 	
	the asymptotic representations of $\Phi(t)$ and $\Psi(t)$, for short times, are  {\it formally} derived by inverting their  Laplace transforms (\ref{eq_3-7}) and (\ref{eq_4-10}),approximated as $s\tau \to \infty$  respectively, see Appendix. 
For large times, the asymptotic representations are promptly obtained by taking the first term of the corresponding Dirichlet series. 
We get:
\begin{equation} \label{eq_5-1}
\qquad \Phi (t) 
 \sim \left \{ 
 \begin{array}{ll}
 {\ds \frac{2}{\sqrt{\pi \tau}} \, t^{-1/2}}, &  t \to 0\,, \\ \\ 
{\ds \frac{4}{\tau} \exp \left( - \lambda _1^2 t / \tau \right)},  &  t\to \infty\,,
\end{array} \right.
\end{equation}
with $\lambda_1^2  \approx 5.78$ and 

\begin{equation} \label{eq_5-2}
\qquad \Psi (t)
\sim \left \{ \begin{array}{ll}
{\ds  \frac{2} { \sqrt{\pi \tau}} \, t^{-1/2}}, &   t \to 0\,, \\ \\
{\ds \frac{8} { \tau}
+ \frac{4}{\tau} \exp \left( - \mu_1^2 t / \tau \right)},  &  t\to \infty\,, 
\end{array}
\right.
\end{equation}
with $\mu_1^2  \approx 26.37$.

  In Figs. 1 and 2 we show  the plots of the relaxation and creep memory functions  versus time, by assuming $\tau=1$ and we compare them
   with their asymptotic representations for short and long times, printed with  dotted and dashed lines, respectively. 
\begin{center}
\includegraphics[width=8cm]{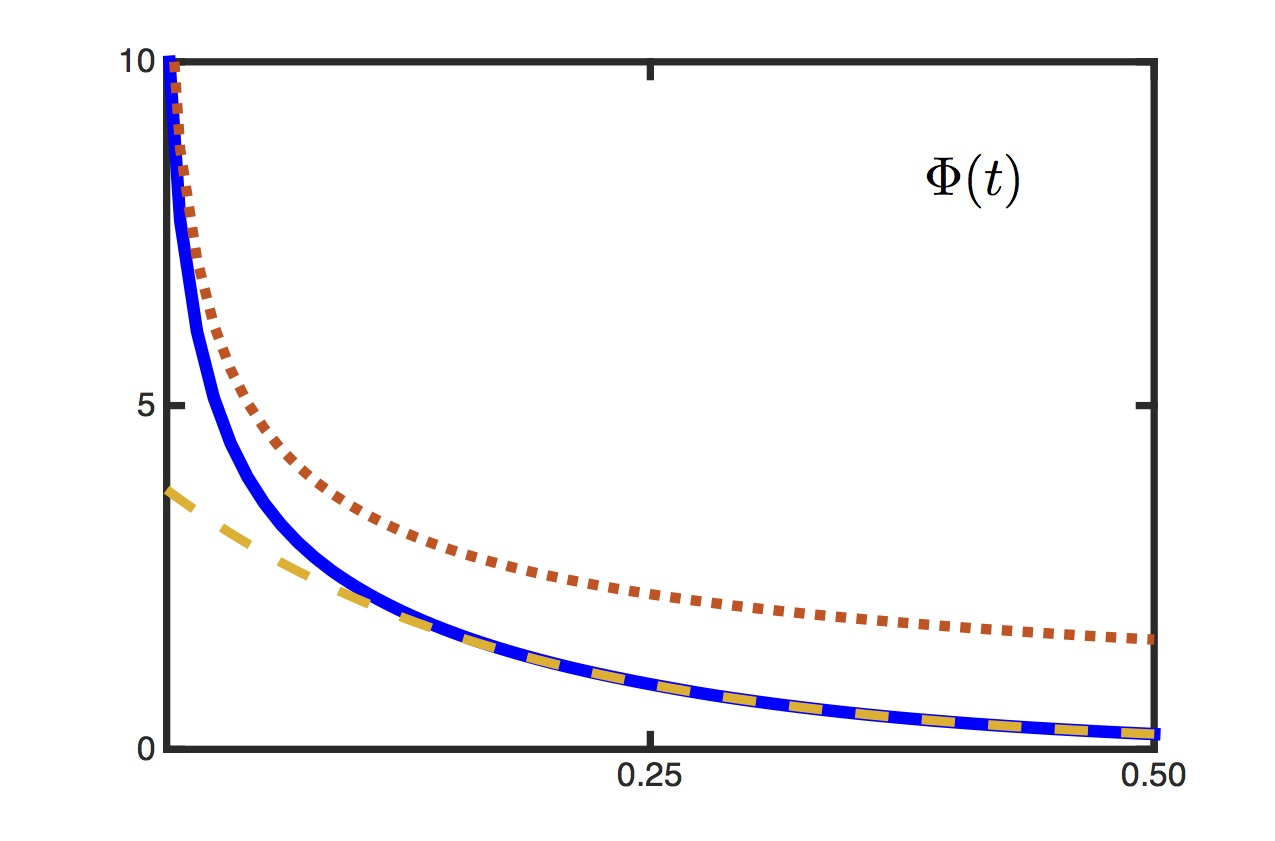}
\end{center}
\vskip -0.5truecm
Fig.1: The relaxation-memory  function $\Phi(t)$ (continuous line) with its asymptotic representations (dotted and dashed lines) versus time scaled with $\tau$. 
\begin{center}
\includegraphics[width=8cm]{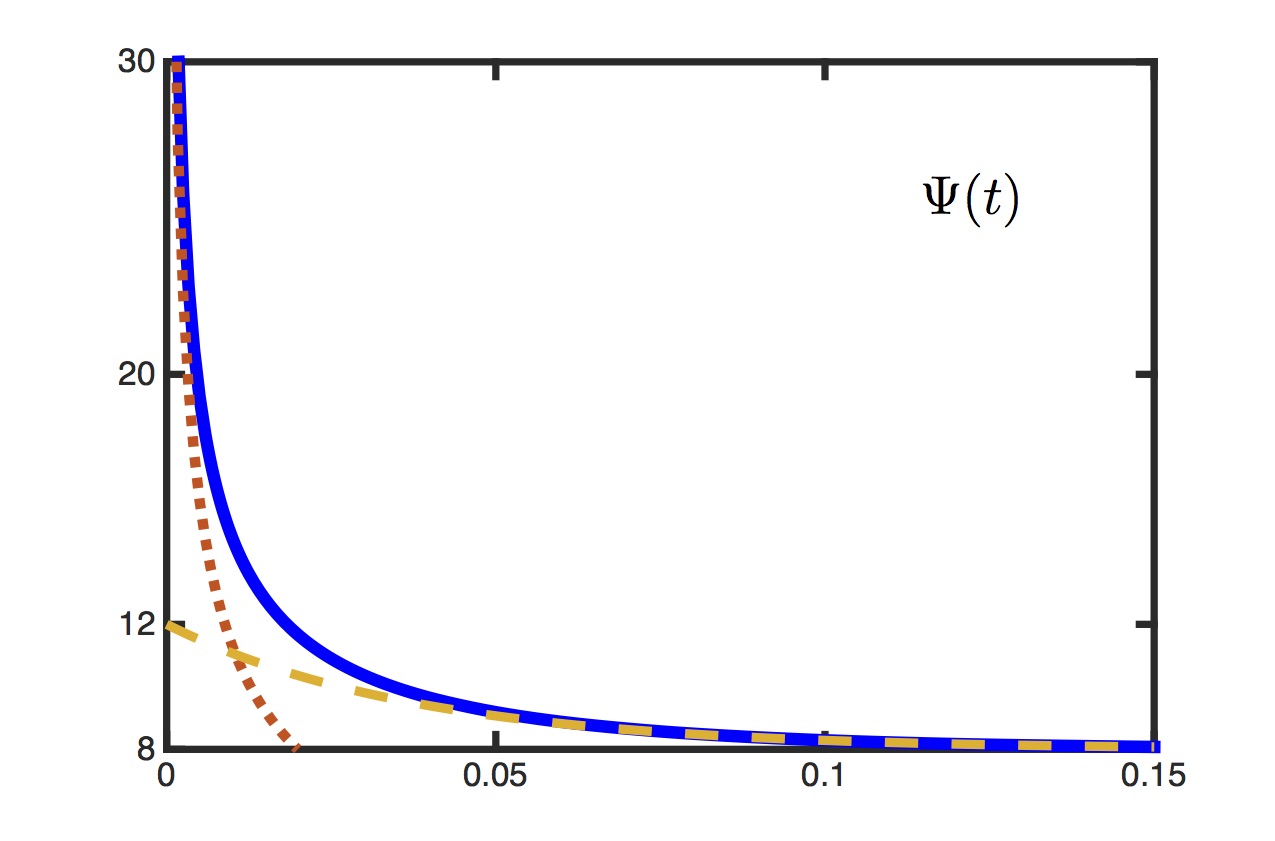}
\end{center}
\vskip -0.5truecm
Fig.2: The creep-memory function $\Psi(t)$ (continuous line) with its asymptotic representations
(dotted and dashed lines)  versus time scaled with $\tau$.

Finally, in Figs. 3 and 4 we show the plots of the $G(t)$ and $J(t)$ respectively, versus time, by assuming $\tau=1$ , $G(0^+) = J(0^+)=1$
in Eqs. (\ref{eq_4-15a}) and (\ref{eq_4-15b}) and taking  100 terms in the corresponding Dirichlet series. 
\begin{center}
\includegraphics[width=8cm]{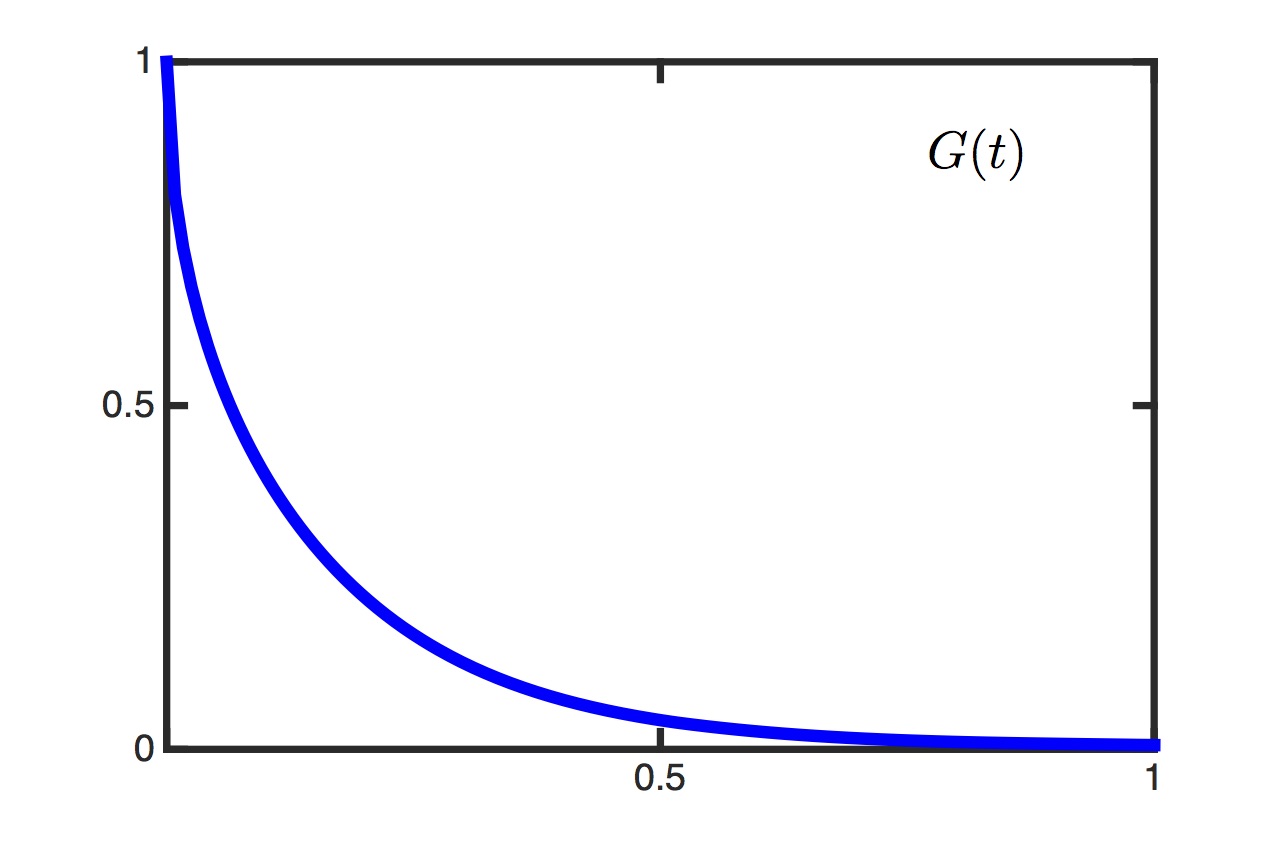}  
\end{center}
\vskip -0.5truecm
Fig.3: The normalized Relaxation Modulus $G(t)$ versus time scaled with $\tau$.


\begin{center}
\includegraphics[width=8cm]{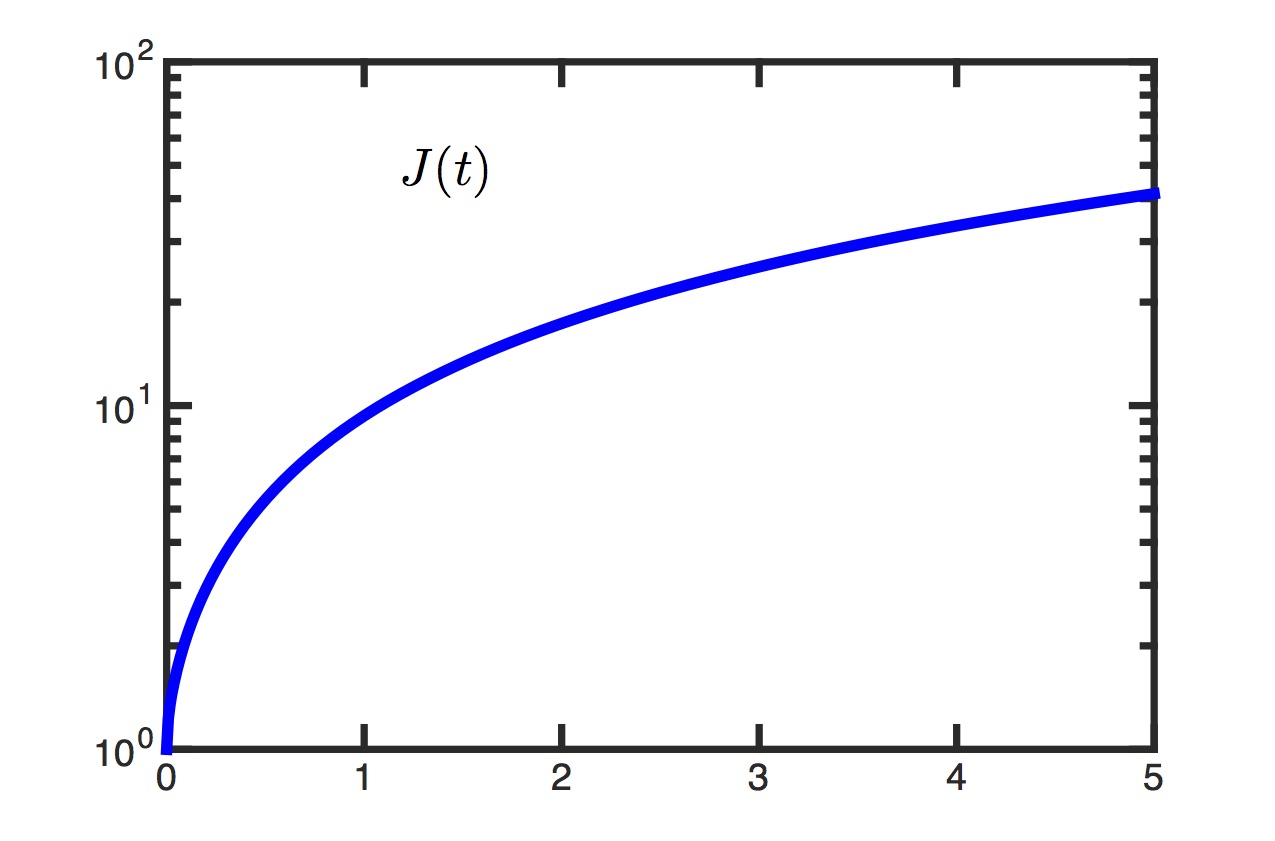}
\end{center}
\vskip -0.5 truecm 
Fig.4: The normalized Creep Compliance $J(t)$ versus time scaled with $\tau$.

\vskip 0.5 cm

Considering the asymptotic and numerical results, as well as the results in \cite{Mainardi 2010} and \cite{Mainardi-Spada 2011}, we observe that the relaxation and creep properties of this model are consistent with those of a peculiar viscoelastic medium that behaves since short times as a fractional Maxwell solid of order 1/2 up to as a standard Maxwell solid for long times. 

\section{Conclusions and final remarks}
	\B{In this paper we have shown that, from the dynamic point of view of uni-axial wave propagation, an elastic tube filled with a viscous  fluid is equivalent to a viscoelastic solid with particular rheological properties that evolve as shown in Section 5.}
We note that this  analogy suffers of subtleties due to the presence of Dirichlet series. 
In any case  this model is worth to be investigated in future with respect 
to the evolution of the transient waves propagating through it.
In fact, as formerly pointed out
by \cite{Buchen-Mainardi 1975}, see also \cite{Mainardi 2010}, 
the singular behavior of the creep function at $t=0$    
induces a wave-front smoothing of any initial discontinuity, like a diffusion effect.
For more details on  this smoothing effect found for transient waves propagating in  media  with  a singular memory we refer to \cite{Hanyga 2001}.   

For that concerning the profile of the fluid inside the tube, we understand that the Poiseuille parabolic profile, corresponding to the Maxwell model, found for long times is preceded by a boundary layer profile, corresponding to the fractional Maxwell model of order 1/2 for short times. This fact is compatible with what expected in a pulsatile blood flow in large arteries for slow and high frequencies  and it is essentially   due to the  effect of the blood viscosity.
For further research we propose to investigate the evolution of the shock formation due to this boundary layer following  the preliminary non-linear analysis by \cite{Mainardi-Buggisch 1983}.  
  
Finally, we note that more realistic models for 1D blood flow take into account the viscoelastic nature of the arterial wall. On this respect the literature is wide but more recent articles require fractional viscoelastic models as well,
 see e.g. \cite{Perdikaris-Karniadakis ABE2014} and references therein.
So we point out the necessity to include in the available 1D models for the arterial  wall viscoelasticity the effects of the blood viscosity discussed in  the present paper.

\vskip 0.5 cm

\noindent \textbf{Acknowledgements}
The work of F. M. has been carried out in the framework of the activities 
of the National Group of Mathematical Physics (GNFM, INdAM) and of
the Interdepartmental Center "L. Galvani"  for integrated studies of Bioinformatics, Biophysics and Biocomplexity of the University of Bologna. The authors are grateful to the anonymous referees for their remarks and suggestions. In particular, we appreciate the comment of one referee who introduced us to the paper in \cite{Sneddon 1960}.


\section*{Appendix: Mathematical discussions}
We provide the details for the proof of some statements found in the text.  
\subsection*{\bf Proof of the expressions (\ref{eq_4-3}) and (\ref{eq_4-11})}
Consider the Laplace Transform of the \textbf{Relaxation-memory function} 
$\widetilde{\Phi} (s)$
\begin{equation} \label{LT_rel}
\widetilde{\Phi} (s ) = \frac{2}{\sqrt{s \tau}} \frac{I_1 (\sqrt{s \tau})}{I_0 (\sqrt{s \tau})}
\end{equation}
where we will consider $\tau = 1$ for sake of simplicity (it can be restored by make the substitution $s \, \leftrightarrow \, s \tau $).
Then, the  Relaxation-memory function is given by
\begin{equation}
\Phi (t) = 4 \sum_{n=1} ^\infty \exp \left( - \lambda _n ^2 t \right) \label{Dirichlet_1}
\end{equation}
where $\lambda _n \in \R$ are such that $J_0 (\lambda _n) = 0 \, , \,\, \forall n = 1, 2, 3, \ldots$ and $t>0$.


\begin{proof}:
Firstly, consider the power series representation for the Modified Bessel functions of the First Kind:
\begin{equation}
\begin{split}
I_{0} (\sqrt{s}) &= 1 + \frac{s}{4} + \frac{s^2}{64} + O(s^3) \, , \\
I_{1} (\sqrt{s}) &= \sqrt{s} \left( \frac{1}{2} + \frac{s}{16} + \frac{s^2}{384} \right) + O(s^{7/2}) \, , \\
\end{split}
\end{equation}
one can eventually deduce that
\begin{equation}
\widetilde{\Phi} (s) = \frac{2}{\sqrt{s}} \frac{I_{1} (\sqrt{s})}{I_{0} (\sqrt{s})} = 2 \,\frac{\frac{1}{2} + \frac{s}{16} + \frac{s^2}{384} + \cdots}{1 + \frac{s}{4} + \frac{s^2}{64} + \cdots}
\end{equation}
which means that the function of our concern is regular in $s=0$ and it does not have any branch cuts.
Then we can obtain the required function by means of the Bromwich Integral:
\begin{equation}
\Phi (t) = \frac{1}{2 \pi i} \int _{Br}  \widetilde{\Phi} (s) \, e^{st} \, ds \, .
\end{equation}
In particular, $\widetilde{\Phi} (s)$ has simple poles such that:
$ I_0 (\sqrt{s}) = 0 $.
Now, if we rename $\sqrt{s}$ as $\sqrt{s} = - i \lambda$, then
$$ I_{0} (\sqrt{s}) = 0 \quad \Longleftrightarrow \quad J_0 (\lambda) = 0 \, . $$
Moreover,
\begin{equation}
\sqrt{s} _n = - i \lambda _n \quad \Leftrightarrow \quad s_n = - \lambda ^2 _n\; , \,\, 
 n \in \N\,.
\end{equation}
From the previous statements, we can then conclude that:
\begin{equation}
\begin{split}
\Phi (t) &= \sum _{s_n} \res \left\{ \widetilde{\Phi} (s) \, e^{st} \right\} _{s_n} =\\
&= \sum _{n=1} ^\infty \res \left\{ \frac{2}{\sqrt{s}} \, \frac{I_1 (\sqrt{s})}{I_0 (\sqrt{s})} \, e^{st} \right\} _{s = - \lambda ^2 _n} \, .
\end{split}
\end{equation}
It is quite straightforward that
\begin{equation}
\begin{split}
\res &\left\{ \frac{2}{\sqrt{s}} \frac{I_{1} (\sqrt{s}) e^{st}}{I_{0} (\sqrt{s})} \right\} _{s = s_n} =\\
&= \lim_{s \to s_n} (s - s_n) \, \frac{2}{\sqrt{s}} \frac{I_{1} (\sqrt{s}) e^{st}}{I_{0} (\sqrt{s})} =\\
&= 4 \exp \left( s_n t \right) \, .
\end{split}
\end{equation}
Thus,
\begin{equation}
\Phi (t) \!=\! \sum _{n=1} ^\infty \res \left\{ \widetilde{\Phi} (s) \, e^{st} \right\} _{s = - \lambda ^2 _n} \!=\! 4 \sum_{n=1} ^\infty e^{- \lambda ^2 _n \, t} \, .
\end{equation}
\end{proof}

Let us now consider the Laplace Transform of the \textbf{Creep-memory function} $\widetilde{\Psi} (s)$
\begin{equation} \label{LT_creep}
\widetilde{\Psi} (s) = \frac{2}{\sqrt{s}} \frac{I_1 (\sqrt{s})}{I_2 (\sqrt{s})} \, .
\end{equation}
Then $\Psi (t)$ is given by
\begin{equation}
\Psi (t) = 8 + 4 \sum_{n=1} ^\infty \exp \left( - \mu ^2 _n \, t \right) \label{Dirichlet_2}
\end{equation}
where $\mu _n$ are such that $J_2 (\mu _n) = 0$ and $\mu _n \neq 0$, for $n \in \N$.

\begin{proof}:
By means of the same procedure shown before 
we are able to point out that
\begin{equation}
\begin{split}
\Psi (t) &= \sum _{s_n} \res \left\{ \widetilde{\Psi} (s) \, e^{st} \right\} _{s_n} =\\
&= \sum _{n = 0} ^\infty \res \left\{ \frac{2}{\sqrt{s}} \, \frac{I_1 (\sqrt{s})}{I_2 (\sqrt{s})} \, e^{st} \right\} _{s = - \mu^2 _n}
\end{split}
\end{equation}
with $\mu_n$ such that $J_2 (\mu _n) = 0$, $\mu _n \neq 0$, f
or $n \in \N$ and $\mu _0 \equiv 0$.\\
Now, we have to distinguish two cases:\\
If $s_n \neq 0$, then
\begin{equation}
\res \left\{ \widetilde{\Psi} (s) \, e^{st} \right\} _{s_n} = 4 \, \exp \left( s_n t \right) \, .
\end{equation}
Otherwise, if $s_n = \mu _0 = 0$,
\begin{equation}
\res \left\{ \widetilde{\Psi} (s) \, e^{st} \right\} _{s_n = 0} = \lim _{s \to 0} s \, \widetilde{\Psi} (s) \, e^{st} = 8 \, .
\end{equation}
Thus,
\begin{equation}
\begin{split}
\Psi (t) &= \sum _{n=0} ^\infty \res \left\{ \widetilde{\Psi} (s) \,
 e^{st} \right\} _{s = - \mu ^2 _n} =\\
 &= 8 + 4 \sum_{n=1} ^\infty \exp \left( - \mu ^2 _n \, t \right) \, .
\end{split}
\end{equation}
\end{proof}

From the above results we are able to conclude that  representation of both 
memory function $\Phi(t)$ and $\Psi(t)$ are given by Dirichlet series (\ref{Dirichlet_1}) and (\ref{Dirichlet_2}), whose convergence is proved in the  following.

\subsection*{{\bf On the convergence of the Dirichlet series (\ref{Dirichlet_1}) and (\ref{Dirichlet_2})}}
The series (\ref{Dirichlet_1}) is convergent for $t>0$.
\begin{proof}:
Consider a Generalized Dirichlet Series:
\begin{equation} \label{Dirichlet_3}
f(z) = \sum _{n=1} ^\infty a_n \, \exp \left(- \alpha _n z \right) \,, \quad \quad z \in \C\,. 
\end{equation}
In general, we have that the abscissa of convergence and the abscissa of absolute convergence would be different, i.e. $\sigma _c \neq \sigma _a$, but they will satisfy the following condition:
\begin{equation}
0 \leq \sigma _a - \sigma _c \leq d = \limsup _{n \to \infty} \frac{\ln n}{\alpha _n} \, .
\end{equation}
If $d=0$, then
\begin{equation}
\sigma \equiv \sigma _c = \sigma _a = \limsup _{n \to \infty} \frac{\ln \left| a_n \right|}{\alpha _n} \, .
\end{equation}
In our case $a_n = 1$ and $\alpha _n = \lambda ^2 _n \neq 0$. Then, we have to understand the behavior of the coefficients $\lambda _n$ for $n \gg 1$, where $J_0 (\lambda _n) = 0 $,  $ \forall n \in \N$.\\
Considering the asymptotic representation
\begin{equation}
J_{0} (x) \overset{x \gg 1}{\sim} \sqrt{\frac{2}{\pi x}} \, \cos \left( x - \frac{\pi}{4} \right)
\end{equation}
we  get to the following conclusion:
\begin{equation}
J_0 (\lambda _n) = 0 \, , \,\,\, \mbox{for} \,\, n \gg 1 \, \Longrightarrow \, \lambda _n \propto n \, , \,\,\, \mbox{for} \,\, n \gg 1 \, .
\end{equation}
Thus,
\begin{equation}
\frac{\ln n}{\alpha _n} = \frac{\ln n}{\lambda ^2 _n} \,\, \overset{n \gg 1}{\sim} \,\, \frac{\ln n}{n^2} \,\, \overset{n \to \infty}{\longrightarrow} \,\, 0
\end{equation}
which tells us that $d=0$.
Finally,
\begin{equation}
\sigma \equiv \sigma _c = \sigma _a = \limsup _{n \to \infty} \frac{\ln \left| a_n \right|}{\alpha _n} = 0
\end{equation}
being $a_n = 1$.\\
This result implies that the series (\ref{Dirichlet_1}), with $a_n = 1$ and $\alpha _n = \lambda ^2 _n \neq 0$, converges for $\texttt{Re} \{ z\} = t > 0$. 
\end{proof}

In a similar way a we can prove that the series (\ref{Dirichlet_2}) is convergent for $t>0$.

\subsection*{{\bf On the asymptotic representations}}
Finally,  we  derive the asymptotic representations for $\Phi (t)$ and $\Psi (t)$
as $t\to 0^+$ 
applying the Tauberian theorems to the corresponding Laplace transforms: 
\begin{equation}
\widetilde{\Phi} (s ) = \frac{2}{\sqrt{s \tau}} \frac{I_1 (\sqrt{s \tau})}{I_0 (\sqrt{s \tau})} \, , 
\quad 
\widetilde{\Psi} (s ) = \frac{2}{\sqrt{s \tau}} \, \frac{I_1 (\sqrt{s \tau})}{I_2 (\sqrt{s \tau})}\,.
\end{equation}
Then,
in view of the asymptotic representation of the modified Bessel functions 
 as  $z\to \infty$ with $|\rm{arg}(z) | <\pi/2$ and for any $\nu$   
$$    I_\nu (z) \sim  \frac{1}{\sqrt{2\pi}} z^{-1/2} \exp (z)\,,$$
see e.g. \cite{AS 1965},
  we  conclude that for $z=\sqrt{s\tau}\to \infty $ ($s>0$) we get
$$    \widetilde \Phi (s)\sim \frac{2}{\sqrt{s\tau}}\,,  \quad 
         \widetilde \Psi (s)\sim \frac{2}{\sqrt{s\tau}}\,, $$
         so that
\begin{equation}
\begin{split}
\Phi (t)&\sim \frac{2}{ \sqrt{\pi \tau}} \, t^{-1/2} \,, \; t \to 0^+\,, \\
\Psi (t)&\sim \frac{2}{ \sqrt{\pi \tau}} \, t^{-1/2} \,, \;  t \to 0^+ \, .
\end{split}
\end{equation}


\begin{thebibliography}{}
%
%

\bibitem[Abramowitz and Stegun(1965)]{AS 1965}
M.~Abramowitz and I.A.~Stegun,
{\it Handbook of Mathematical Functions}, Dover, New York (1965).

\bibitem[Barclay, Moodie and Madan(2004)] {BMM 1981}
D.W. ~Barclay, T.B. Moodie and V.P. Madan,
Linear and non-linear pulse propagation in fluid-filled compliant tubes,
     {\em Meccanica} {16},~3--9 (2004).  

\bibitem[Barnard et al.(1966)]{Varley-et-al 1966}
A.C.L.~Barnard, W.A. ~Hunt, W.P.~Timlake, E.~Varley,
A theory of  fluid flow in compliant tubes,
   {\em Biophys. J.} {6}, 717--724 (1966).  

\bibitem[Buchen and Mainardi(1975)]{Buchen-Mainardi 1975}
P.W. Buchen, F. Mainardi,
Asymptotic expansions for transient viscoelastic waves,
{\it Journal de M{\'e}canique} {14},  597--608 (1975).


\bibitem[Daidzic(2014)]{Daidzic JFE2014}
N.E. Daidzic,
Application of Womersley model to reconstruct  pulsatile flow
from Doppler ultrasound measurements,
{\it Journal of Fluid Engineering} {136}, 041102/1-15 (2014).

\bibitem[Giusti and Mainardi (2014)]{AG-FM-ICMMB-2014}
A. Giusti, F. Mainardi, 
A linear viscoelastic model for arterial pulse propagation, 
{\em Proceedings of ICMMB-19}, 168--171 (2014).

\bibitem[Giusti and Mainardi (2016)]{AG-FM-eprint}
A. Giusti, F. Mainardi, 
On infinite series concerning zeros of Bessel functions of the first kind, 
{\em e-print}, \href{http://arxiv.org/abs/1601.00563}{arXiv:1601.00563} (2016).

\bibitem[Gorenflo et al.(2014)]{Getal-2014}
R. ~Gorenflo, A. ~Kilbas, F. ~Mainardi, S. ~Rogosin,
{\it Mittag-Leffler Functions, Related Topics and Applications},
Springer, New York (2014).

\bibitem[Hanyga (2001)]{Hanyga 2001}
A. Hanyga, 
Wave propagation in media with singular memory,
{\it Mathematical and Computer Modelling} {34},  1399--1421 (2001).

\bibitem[Hardy and Riesz(1915)]{Hardy-Riesz 1915}
G.H. Hardy, M. Riesz, 
{\it The General Theory of Dirichlet Series},
Cambridge University Press, Cambridge (1915).

\bibitem[He et al.(1992)]{He-et-al ABE1993}
X. He, D.N. Ku, J.E. Moors Jr,
Simple calculation of the velocity profiles for pulsatile flow
in a blood vessel using Mathematica,
{\it Annals of Biomedical Engineering} {21},  45--49 (1993).

\bibitem[Hoogstraten and Smit(1978)]{Hoogstraten-Smit 1978}
H.W.~Hoogstraten, G.H.~Smit, 
 A mathematical theory of shock-wave formation in arterial blood flow,
{\em Acta Mech.}  {\bf 30},  145--155 (1978).  

\bibitem[Koeller(2010)]{K-2010}
   R.C. ~Koeller,
A theory relating creep and relaxation for linear materials with memory,
{\it J. Appl. Mech.} 77, 031008-1-031008-9 (2010).

\bibitem[Mainardi(2010)]{Mainardi 2010}
 F.~Mainardi, 
     {\it Fractional Calculus and Waves in Linear Viscoelasticity},
     Imperial College Press,  London (2010). 

\bibitem[Mainardi and Buggisch(1983)]{Mainardi-Buggisch 1983}
F.~Mainardi, H.~Buggisch,
On non-linear waves in liquid-filled elastic tubes,
in U. Nigul and J. Engelbrecht (Eds), {\it Nonlinear Deformation Waves},
Springer, Berlin, 1983, pp.~87--100.

\bibitem[Mainardi and Spada(2011)]{Mainardi-Spada 2011}
F. ~Mainardi, G. ~Spada,
Creep, relaxation and viscosity properties for basic fractional models in rheology,
{\it Eur. Phys. J. Special Topics} {193}, 133--160 (2011).
[E-print http://arxiv.org/abs/1110.3400]

\bibitem[Perdikaris and Karniadakis(2014)]{Perdikaris-Karniadakis ABE2014}
P. Perdikaris, G.E. Karniadakis,
Fractional-order viscoelasticity in one-dimensional blood flow  models,
{\it Ann. Biomed. Eng.} Vol 42 No 5 1012--1023 (2014)
[DOI: 10.1007/s10439-014-0970-3]

\bibitem[Rabotnov(1980)]{Rabotnov-1980}
Yu.N. Rabotnov,
{\it Elements of Hereditary Solid Mechanics}, 
Mir Publishers, Moscow (1980).

\bibitem[Rogosin and Mainardi(2014)]{MR-2014}
S. Rogosin, F. Mainardi,  George William Scott Blair - the pioneer of factional calculus in rheology,
{\it Commun. Appl. Ind. Math.} {6}, pp. 20 (2014) DOI: 10.1685/journal.caim.481

\bibitem[Sneddon(1960)]{Sneddon 1960}
I. N. Sneddon,
On some infinite series involving the zeros of Bessel functions of 
the first kind,
{\it Proc. Glasgow Math. Assoc.} {4}, 144-156 (1960).

\bibitem[Womersley(1955)]{Womersley 1955}
   J.R. ~Womersley,
Method for the calculation of velocity, rate of flow and viscous drag in arteries when the pressure gradient is known,
{\it J. Physiol.} {127},  553--563 (1955). 
 
\bibitem[Womersley(1957)]{W-1957}
   J.R. ~Womersley,
{\it An Elastic Tube Theory of Pulse Transmission and Oscillatory
Flow in Mammalian Arteries},
Wright Air Development Center, Technical Report, WADC-TR,  56-614 (1957). 

\end{thebibliography}


\end{document}